\documentstyle[epsfig,aps,pre]{revtex}
\tightenlines
\begin{document}
\draft
\title{Heat and momentum transport in a multicomponent mixture 
far from equilibrium}
\author{V. Garz\'o\footnote[1]{Electronic address: vicenteg@unex.es}}
\address{Departamento de F\'{\i}sica, Universidad de Extremadura,\\
E-06071 Badajoz, Spain}
\date{\today}
\maketitle
\begin {abstract}
Explicit expressions for the heat and momentum fluxes are given for a 
low-density multicomponent mixture in a steady state with temperature and 
velocity gradients. The results are obtained from a formally exact solution 
of the Gross-Krook model [Phys. Rev. {\bf 102}, 593 (1956)] of the Boltzmann 
equation for a multicomponent mixture. The 
transport coefficients (shear viscosity, viscometric functions, thermal 
conductivity and a cross coefficient measuring the heat flux orthogonal to 
the thermal gradient) are nonlinear functions of the velocity and 
temperature gradients and the parameters of the mixture (particle masses, 
concentrations, and force constants). The description applies for conditions 
arbitrarily far from equilibrium and is not restricted to any range of mass 
ratios, molar fractions and/or size ratios.
The results show that, in general, the presence of the shear flow produces 
an inhibition in the transport of momentum and energy with respect to 
that of the Navier-Stokes regime. In the particular case of 
particles mechanically equivalent and in the tracer limit, previous results 
are recovered. \\

Keywords: Couette flow; Gross-Krook kinetic model; 
Multicomponent mixture; Nonlinear transport.

\end {abstract}
\pacs{PACS number(s) : 51.10.+y, 05.20.Dd, 05.60.+w, 47.50.+d}

\section{Introduction}
\label{sec_1}

Needless to say, the study of transport phenomena in fluid mixtures is 
much more complicated than that of a single fluid. Not only is the number 
of transport coefficients much higher but also they depend on parameters 
such as the mass ratios, the molar fractions and/or the size ratios. These 
difficulties increase considerably when one attempts to analyze far from 
equilibrium states for which the linear relationships between 
fluxes and gradients do not apply. Due to the the complexity of the general 
problem, tractable specific situations must be considered. The study
of such situations (for which a complete description can be given) is of 
great value since they allow us to gain some insight into the understanding 
of more general problems.

In order to capture the relevant aspects of nonlinear transport phenomena, 
a low-density $N$-component mixture with short-range interactions can be 
chosen as a prototype system. In this case, the essential information on 
the mixture is given by the one-particle velocity distribution functions 
$f_i({\bf r},{\bf v};t)$ ($i= 1,\ldots,N$), which obey the set of coupled
nonlinear Boltzmann equations \cite{M89}. For states near equilibrium, the 
constitutive Navier-Stokes equations for the heat and momentum fluxes 
(Fourier's heat law and Newton's viscosity law) can be derived, for 
instance, by using the Chapman-Enskog method \cite{CC70}. Nevertheless, 
beyond the Navier-Stokes domain, the Chapman-Enskog calculations are 
prohibitively difficult and not applicable for large gradients. An 
interesting problem in which some insight into the mechanisms of heat and 
momentum transport far from equilibrium can be gained is that of steady 
planar Couette flow. The physical situation of the system is that of a 
multicomponent gas enclosed between two parallel plates in relative motion 
and maintained, in general, at different temperatures. These boundary 
conditions lead to temperature and density gradients coexisting with a 
velocity field. In the steady Couette flow there are two local parameters 
measuring the departure from equilibrium: the shear rate and the thermal 
gradient. Our aim is to get the relevant transport coefficients in terms of 
both gradients, as well as in terms of the parameters characterizing the 
mixture.

Unfortunatelly, due to the mathematical difficulties embodied in the {\em 
nonlinear} Couette flow problem, no analytic solution to the Boltzmann 
equation valid for {\em arbitrary} values of the shear rate and the 
thermal gradient is known, even for a single gas. Only in the limit of small 
shear rates, a perturbation solution in powers of the shear field through 
super-Burnett order has been recently worked out for a single gas of Maxwell 
molecules \cite{TS95}. If one wants to obtain the full nonlinear dependence 
of the transport coefficients on the imposed gradients, either one performs 
computer simulations or on the analytical side one considers kinetic models 
or the Grad method \cite{RC97}. Here, we take the second 
route and consider a kinetic model. Specifically, we use the nonlinear 
kinetic model for mixtures proposed by Gross and Krook (GK) \cite{GK56}. 
This model is constructed in the same spirit as the well-known 
Bhatnagar-Gross-Krook (BGK) model of a single gas \cite{BGK54}, which admits 
an exact solution for the steady nonlinear Couette flow \cite{BSD87,KDSB89}. 
The GK model preserves the most important physical properties of the true 
Boltzmann equation while allowing for more complete calculations.

As mentioned above, the goal of this paper is to obtain exactly the 
pressure tensor (momentum transport) and the heat flux (energy transport) of 
a multicomponent mixture under Couette flow in the context of the GK model. 
We assume that the molar fractions are constant, so that no mutual diffusion 
appears in the system. On the other hand, no restriction on the masses, 
concentrations, and force constants will be considered. Consequently, the 
transport coefficients are {\em nonlinear} functions of both the shear 
rate and the thermal gradient and of the parameters of the mixture. 
Obviously, our results reduce to those obtained from the BGK model in the 
case of particles mechanically equivalent \cite{BSD87}. Further, when one 
considers a binary mixture and takes the tracer limit (molar fraction of one 
of the components much smaller than $1$), previous results derived from the 
GK model are also recovered \cite{GS93}.

The plan of the paper is as follows. Section \ \ref{sec2} starts with a 
brief survey of the GK model for an $N$-component mixture. Then, the 
nolinear Couette flow is introduced. In the same way as in the single gas 
case, we propose a solution characterized by constant pressure and linear 
velocity and parabolic temperature profiles with respect to a scaled 
variable. The consistency of the assumed profiles is fulfilled in Sec.\ 
\ref{sec3} from a formal solution of the GK model. The consistency 
condition for the temperature yields a system of $N$ implicit equations 
which give the shear rate dependence of the partial temperatures $T_i$ of 
each species. For nonzero shear rates, the $T_i$'s are different, so that 
the kinetic energy is not equally distributed among the different species. 
This effect is generic for multicomponent systems in far from 
equilibrium states and is consistent with other studies of mixtures out of 
equilibrium \cite{MGS94,MGS95,MMG96}. Apart from the partial temperatures, 
the pressure tensor and the heat flux vector are also explicitly determined. 
From these fluxes the main transport coefficients of the problem can be 
identified. Specifically, there are five relevant transport coefficients: 
the generalized shear viscosity function, two viscometric functions, the 
generalized thermal conductivity function, and a coefficient measuring 
cross effects in the heat flux (heat flux along the $x$ direction due to a 
thermal gradient along the $y$ direction). The expressions of these 
functions are written in Sec.\ \ref{sec4}. The particular case of a binary 
mixture ($N=2$) is considered in Sec.\ \ref{sec5}, where the 
shear rate dependence of the above transport coefficients is illustrated for 
several values of the parameters of the mixture. Finally, we close the paper 
in Sec.\ \ref{sec6} with a brief discussion.

\section{The kinetic model and the problem}
\label{sec2}

We consider an $N$-component mixture. In the low-density regime, the 
one-particle velocity distribution function $f_i({\bf r},{\bf v};t)$
of species $i$ ($i=1,\ldots,N$) obeys the Boltzmann equation \cite{CC70} 
\begin{equation}
\label{2.1}
\frac{\partial}{\partial t}f_i+{\bf v}\cdot \nabla f_i=K_{ii}[f_i,f_i]+
\sum_{j\neq i}K_{ij}[f_i,f_j],
\end{equation}
where $K_{ij}[f_i,f_j]$ is the nonlinear Boltzmann collision operator. The 
first and second terms on the right-hand side represent self- and 
cross-collisions, respectively. They conserve the number of particles of 
each species, the total momentum and the total energy. The local number 
density $n_i$ and mean velocity ${\bf u}_i$ of species $i$ are defined as
\begin{equation}
\label{2.2}
n_i=\int d{\bf v} f_i\;,
\end{equation}
\begin{equation}
\label{2.3}
{\bf u}_i=\frac{1}{n_i}\int d{\bf v} {\bf v} f_i \;.
\end{equation}
These quantities define the total number density $n=\sum_i n_i$ and 
the flow velocity  
 \begin{equation}
\label{2.4}
{\bf u}=\frac{1}{\rho}\sum_{i=1}^{N} \rho_i {\bf u}_i,
\end{equation}
where $\rho_i=m_in_i$ and $\rho=\sum_i \rho_i$, $m_i$ being the mass of a 
particle of species $i$. It is also convenient to define a temperature 
$T_i$ for each species through 
\begin{equation}
\label{2.5}
\frac{3}{2}n_ik_BT_i=\frac{m_i}{2} \int d{\bf v} ({\bf v}-{\bf u}_i)^2 f_i
\;,
\end{equation}
which is a measure of its mean kinetic energy per particle. Here, $k_B$ is 
the Boltzmann constant. From these partial temperatures, the temperature 
$T$ of the mixture (which is the relevant one at a hydrodynamic level) is 
given by 
\begin{equation}
\label{2.6}
nk_BT=\sum_{i=1}^{N}\left[ n_ik_BT_i+\frac{1}{3}\rho_i({\bf u}_i-{\bf 
u})^2\right].
\end{equation}

The corresponding balance equations associated with $n_i$, ${\bf u}$ and 
$T$ can be easily obtained from the set of Boltzmann equations (\ref{2.1}):
\begin{equation}
\label{2.n1}
D_tn_i+n_i\nabla \cdot {\bf u}+\frac{\nabla \cdot {\bf J}_i}{m_i}=0,
\end{equation}
\begin{equation}
\label{2.n2}
D_t {\bf u}+\rho^{-1}\nabla {\sf P}=0,
\end{equation} 
\begin{equation}
\label{2.n3}
D_tT-\frac{T}{n}\sum_i \frac{\nabla \cdot {\bf J}_i}{m_i}+\frac{2}{3nk_BT}
\left(\nabla\cdot {\bf q}+{\sf P}:\nabla {\bf u}\right)=0.
\end{equation}
In the above equations, $D_t\equiv\partial_t+{\bf u}\cdot \nabla$, and we 
have introduced the dissipative fluxes of mass
\begin{equation}
\label{2.n4}
{\bf J}_i=m_i\int d{\bf v} {\bf V} f_i,
\end{equation}
momentum (pressure tensor)
\begin{equation}
\label{2.n5}
{\sf P}=\sum_{i=1}^{N}m_i\int d{\bf v} {\bf V} {\bf V} f_i=\sum_{i=1}^{N}
{\sf P}_i,
\end{equation}
and energy (heat flux)
\begin{equation}
\label{2.n6}
{\bf q}=\sum_{i=1}^{N}\frac{m_i}{2}\int d{\bf v} V^2 {\bf V} 
f_i=\sum_{i=1}^{N}
{\bf q}_i,
\end{equation}
where ${\bf V}={\bf v}-{\bf u}$ is the peculiar velocity. These fluxes 
define the relevant transport coefficients of the mixture.

In general, the set of $N$ nonlinear 
Boltzmann equations (\ref{2.1}) cannot be solved, especially in far from 
equilibrium situations. This is basically due to the complex mathematical 
structure of the Boltzmann collision operators. In order to overcome such 
difficulties one can resort to kinetic models in which case the Boltzmann 
collision operators are replaced by simpler terms that preserve their main 
physical properties. Here we consider the well-known Gross-Krook (GK) model 
\cite{GK56}, where the Boltzmann terms $K_{ij}[f_i,f_j]$ in Eqs.\ 
(\ref{2.1}) are replaced by simple relaxation terms of the form 
\begin{equation}
\label{2.7}
K_{ij}^{\text{GK}}=-\nu_{ij}(f_i-f_{ij})\;,
\end{equation} 
where $\nu_{ij}$ is an effective collision frequency and 
the reference distribution functions $f_{ij}$ are given by 
\begin{equation}
\label{2.8}
f_{ij}=n_i\left(\frac{m_i}{2\pi k_BT_{ij}}\right)^{3/2} \exp
 \left[-\frac{m_i}{2k_BT_{ij}}({\bf v}-{\bf u}_{ij})^2\right]\;.
\end{equation}
Here, we have introduced the fields
\begin{equation}
\label{2.9}
{\bf u}_{ij}=\frac{m_i{\bf u}_i+m_j{\bf u}_j}{m_i+m_j} \;,
\end{equation}
\begin{equation}
\label{2.10}
T_{ij}=T_i+2\frac{m_im_j}{(m_i+m_j)^2}\left[(T_j-T_i)+\frac{m_i}{6k_B}
({\bf u}_i-{\bf u}_j)^2\right]\;.
\end{equation}
The above expressions are obtained by requiring that momentum and energy 
moments of $K_{ij}^{\text{GK}}$ be the same as those of the Boltzmann 
operator for Maxwell molecules (i.e., particles interacting through a 
potential of the form $\Phi_{ij}=\kappa_{ij}r^{-4}$). This 
allows one to identify $\nu_{ij}$ as \cite{GK56} 
\begin{equation}
\label{2.11}
\nu_{ij}=A n_j \left[ \kappa_{ij} \frac{m_i+m_j}{m_im_j} 
\right] ^{1/2} \;,
\end{equation}
where $A=4\pi\times0.422$. Although we are considering Maxwell molecules, we 
can assign an effective diameter $\sigma_{ij}$ to the interaction between 
particles of species $i$ and $j$. Dimensional analysis allows one to 
interpret the ratio $(\kappa_{ij}/\kappa_{jj})^{1/4}$ as the size ratio 
$\sigma_{ij}/\sigma_{jj}$. On the other hand, it must be remarked that the 
results derived in this paper could in principle be extended to more general 
potentials.

We now describe the problem we are interested in. Let us assume that the 
$N$-component mixture is enclosed between two parallel plates in relative 
motion and kept at different temperatures. Let the $x$-axis be parallel to 
the direction of motion and the $y$-axis be normal to the plates. We want to 
analyze a steady state with velocity ($u_x$) and temperature ($T$) gradients 
along the $y$ direction (steady Couette flow). 
These boundary conditions give rise to heat and momentum transport, the 
relevant fluxes being the pressure tensor and the heat flux. Our goal is to 
get these transport properties in terms of the imposed gradients and the 
parameters of the mixture, namely, masses, concentrations, and ``sizes'' of 
the particles. In the geometry of the steady Couette flow and in the context 
of the GK model, the set of $N$ coupled  equations for the mixture reads 
\begin{eqnarray}
\label{2.12}
v_y\frac{\partial}{\partial y}f_i&=&
-\nu_{ii}\left( f_i-f_{ii}\right)-\sum_{j\neq i}\nu_{ij}\left( f_i-f_{ij}
\right)\nonumber\\
&=&-\nu_if_i+\sum_{j=1}^{N}\nu_{ij} f_{ij},
\end{eqnarray} 
where 
\begin{equation}
\label{2.13}
\nu_i=\sum_{j=1}^{N} \nu_{ij}=\nu_{ii}+\sum_{j\neq i}\nu_{ij}
\end{equation}
is the total collision frequency for a particle of species $i$. In the same 
way as in previous works, here we are interested in studying the 
properties of the mixture in the bulk region far away from 
the plates. Thus, instead of introducing the 
appropriate boundary conditions for Eqs.\ (\ref{2.12}), we guess the 
spatial dependence of the hydrodynamic fields and then we verify their 
consistency. In this sense, we expect to describe transport phenomena in the 
bulk of the system by looking for a consistent solution regardless of the 
actual features of the boundaries. 

In the case of all the species being mechanically identical, 
${\bf u}_i={\bf u}$, $T_i=T$ and the GK model reduces to the well-known 
(closed) BGK equation \cite{BGK54} for the total distribution function 
$f=\sum_{i}f_i$. In this case, the BGK model admits a consistent solution 
characterized by a constant pressure, and linear velocity $u_x$ and 
parabolic temperature $T$ profiles with respect to a scaled variable 
\cite{BSD87,KDSB89}. This variable takes into account the fact that the rate 
at which collisions take place is nonuniform. The above solution is not 
restricted to small velocity and temperature gradients and thus provides a 
rare detailed description of nonlinear heat and momentum transport far from 
equilibrium. Guided by these results for the one-component case, a similar 
solution is proposed here for the mixture. First, on physical 
grounds one expects that the steady Couette flow for the mixture is 
characterized by (a) constant molar fractions $x_i=n_i/n$, (b) the absence 
of mutual diffusion, i.e., ${\bf u}_i={\bf u}$, and (c) constant temperature 
ratios $\chi_i\equiv T_i/T$. In addition, we assume that the partial 
pressure $p_i=n_ik_BT_i$, the velocity $u_{i,x}$, and the partial 
temperature $T_i$ have the forms:
\begin{equation}
\label{2.14}
p_i\equiv n_ik_BT_i=\mbox{\rm const}\;,
\end{equation}
\begin{equation}
\label{2.15}
\frac{1}{\nu_{i}(y)}\frac{\partial}{\partial y} u_{i,x}=a_i=
\mbox{\rm const}\;, 
\end{equation}
\begin{equation}
\label{2.16}
\left[ \frac{1}{\nu_{i}(y)}\frac{\partial}{\partial y}\right]^2
T_i=-\frac{2m_i}{k_B} \gamma_i(a_i) =\mbox{\rm const}\;.
\end{equation} 
The dimensionless functions $\gamma_i$ must be explicitly determined by 
consistency. The conditions ${\bf u}_i={\bf u}$ and $\chi_i=\mbox{\rm 
const}$ imply, respectively, that 
\begin{equation}
\label{2.16bis}
\nu_ia_i=\nu_ja_j,
\end{equation} 
\begin{equation}
\label{2.17}
\frac{m_i\gamma_i\nu_i^2}{\chi_i}=\frac{m_j\gamma_j\nu_j^2}{\chi_j},
\end{equation} 
where use has been made of the fact that the ratios $\nu_i/\nu_j$ are 
constant. According to Eqs.\ (\ref{2.16bis}) and (\ref{2.17}), only one 
of the shear rates $a_i$ and only one of the parameters $\gamma_i$ 
($i=1,\ldots,N$) is independent. The temperature ratios $\chi_i$ 
depend on the shear rate and on the parameters of the mixture. Obviously, 
$\chi_i=1$ if (i) the mixture is at equilibrium or (ii) the particles are 
mechanically equivalent ($m_i=m$ and $\kappa_{ij}=\kappa$). The explicit 
calculation of the temperature ratios is one of the main objectives of this 
paper.

To verify the consistency of the assumed solution it is necessary to prove 
that 
\begin{equation}
\label{2.18}
\int d{\bf v} \left\{1,{\bf 
V},V^2\right\}f_i=\left\{n_i,0,3\frac{p_i}{m_i}\right\}.
\end{equation} 
The fulfillment of these conditions will be carried out in the next Section.

\section{Consistency of the solution. Calculation of the fluxes}
\label{sec3}

In this Section we show that the relations (\ref{2.18}) are 
verified and obtain the pressure tensor and the heat fux. To do 
that, let us consider the formal solution to Eq.\ (\ref{2.12}) given by  
\begin{equation}
\label{3.1}
f_i=\left(1+\frac{v_y}{\nu_i}\frac{\partial}{\partial y}\right)^{-1}
\sum_{j=1}^{N}\frac{\nu_{ij}}{\nu_i}f_{ij}.
\end{equation}
Notice that when the operator $[1+(v_y/\nu_i)(\partial/\partial y)]^{-1}$ 
acts on the hydrodynamic quantities the expressions 
(\ref{2.14})--(\ref{2.16}) are assumed to hold. By using this formal 
solution in the Appendix we show that the conditions for the density and the 
flow velocity are satisfied with independence of the value of the parameters 
$\gamma_i$ in Eq.\ (\ref{2.16}) for the temperature $T_i(y)$. A similar 
analysis for the temperature condition shows that a solution exists only for 
a particular choice of $\gamma_i$ given implicitly by the equation:
\begin{equation}
\label{3.2} 
\frac{3}{2}\sum_{j\neq i}\left(1-\chi_{ij}\right) \nu_{ij}=\sum_{j=1}^{N}
\nu_{ij} \chi_{ij}\left\{ a_i^2 F_1(\gamma_{ij})-
\gamma_{ij}\left[2F_2(\gamma_{ij})+3
F_1(\gamma_{ij})\right]\right\}. 
\end{equation} 
Here, we have introduced the 
quantities $\gamma_{ij}=\chi_{ij}\gamma_i$, 
\begin{equation}
\label{3.3}
\chi_{ij}=1+2\frac{m_im_j}{(m_i+m_j)^2}\left(\frac{\chi_j}{\chi_i}-1\right).
\end{equation}
In addition, the functions $F_r(\gamma_{ij})$ are defined as 
\begin{equation}
\label{3.3bis}
F_r(\gamma_{ij})=\left(\frac{d}{d
\gamma_{ij}}\gamma_{ij}\right)^rF_0(
\gamma_{ij}) 
\end{equation}
with
\begin{equation}
\label{3.4}
F_0(\gamma_{ij})=\frac{2}{\gamma_{ij}}
\int_0^{\infty} dt \, t \, e^{-t^2/2} 
K_0(2\gamma_{ij}^{-1/4}t^{1/2})\;,
\end{equation}
$K_0$ being the zeroth-order modified Bessel function \cite{Abramowitz}. 
The set of $N$ coupled equations (\ref{3.2}) must be numerically solved 
subject to the constraints (\ref{2.17}) and 
\begin{equation}
\label{3.5}
\sum_{i=1}^{N} x_i \chi_i=1 .
\end{equation}
From a mathematical point of view, the problem is now well 
posed since we have $N$ independent equations (\ref{3.2})
for a set of $N$ unknowns, say for instance
$\{\gamma_1,\chi_1,\chi_2,\cdots,\chi_{N-1}\}$. The solution of these $N$ 
implicit equations gives the above quantities as functions of the shear rate 
and the parameters of the mixture (masses, concentrations, and sizes). 
This completes the confirmation of the consistency conditions for the 
hydrodynamic fields.

Once the functions $\gamma_i$ and $\chi_i$ are known, one can derive 
explicit expressions for the fluxes. In the case of the Couette flow, the 
mass flux ${\bf J}_i$ vanishes, and so the relevant fluxes are the pressure 
tensor (\ref{2.n5}) and the heat flux (\ref{2.n6}). We begin with the 
nonzero elements of the pressure tensor. In the Appendix it is shown that 
they can be written as
\begin{equation}
\label{3.6} 
P_{xx}=nk_BT\sum_{i=1}^{N}\frac{x_i\chi_i}{\nu_i}\sum_{j=1}^{N}\nu_{ij}
\chi_{ij}\left\{1+4\gamma_{ij}\left[F_1(\gamma_{ij})+
F_2(\gamma_{ij})\right]\right\},
\end{equation} 
\begin{equation}
\label{3.7} 
P_{yy}=nk_BT\sum_{i=1}^{N}\frac{x_i\chi_i}{\nu_i}\sum_{j=1}^{N}\nu_{ij}
\chi_{ij}\left\{1-2\gamma_{ij}\left[F_1(\gamma_{ij})+
2F_2(\gamma_{ij})\right]\right\},
\end{equation} 
\begin{equation}
\label{3.8} 
P_{zz}=nk_BT\sum_{i=1}^{N}\frac{x_i\chi_i}{\nu_i}\sum_{j=1}^{N}\nu_{ij}
\chi_{ij}\left[1-2\gamma_{ij}F_1(\gamma_{ij})\right],
\end{equation}  
\begin{equation}
\label{3.9} 
P_{xy}=-nk_BT\sum_{i=1}^{N}\frac{x_i\chi_i}{\nu_i^2}\sum_{j=1}^{N}\nu_{ij}
\chi_{ij}F_0(\gamma_{ij}) \frac{\partial u_x}{\partial y}.
\end{equation}  
Notice that the pressure tensor does not depend on the thermal 
gradient. With respect to the heat flux, its nonzero components are $q_x$ 
and $q_y$. The fact that $q_x\neq 0$ is a direct consequence of the presence 
of the shear flow. These components are also evaluated in the Appendix and 
the result is 
\begin{eqnarray}
\label{3.10}
q_y&=&-\frac{5}{2}nk_B^2T 
\sum_{i=1}^{N}\frac{x_i\chi_i^2}{5m_i\nu_i^2}\sum_{j=1}^{N}\nu_{ij}
\chi_{ij}\left\{3F_1(\gamma_{ij})+2F_2(\gamma_{ij})
\right. \nonumber\\
& & \left.
+2a_i^2\left[4F_4(\gamma_{ij})+4F_3(\gamma_{ij})+
F_2(\gamma_{ij})\right]\right\}\frac{\partial T}{\partial y},
\end{eqnarray}
\begin{eqnarray}
\label{3.11}
q_x&=&nk_B^2T 
\sum_{i=1}^{N}a_i\frac{x_i\chi_i^2}{m_i\nu_i^2}\sum_{j=1}^{N}\nu_{ij}
\chi_{ij}\left\{5F_2(\gamma_{ij})+2F_3(\gamma_{ij})
\right.\nonumber\\
& & \left.
+2a_i^2\left[4F_5(\gamma_{ij})+
8F_4(\gamma_{ij})+5F_3(\gamma_{ij})+
F_2(\gamma_{ij})\right]\right\}\frac{\partial T}{\partial y}.
\end{eqnarray}
From Eqs.\ (\ref{3.9}) and (\ref{3.10}), it is easy to prove the 
identity $\partial_y q_y=-P_{xy} \partial_y u_x$, which is in fact required 
by the energy conservation equation (\ref{2.n3}) to support the forms for 
the hydrodynamic profiles.

\section{Generalized transport coefficients}
\label{sec4}

It is useful to summarize the above results for 
the fluxes in terms of appropriate scalar transport coefficients. To do 
that, let us reduce the velocity gradient 
$\partial_y u_x$ with respect to an effective collision frequency $\nu$. 
Since our description applies for arbitrary mass, concentration and size 
ratios, we choose for simplicity the average collision frequency 
\begin{equation}
\label{3.12}
\nu=\frac{1}{N(N-1)}\sum_{i=1}^{N}\sum_{j\neq i}\frac{n}{n_j}\nu_{ij}
\end{equation}
and define the (dimensionless) shear rate 
\begin{equation}
\label{3.12bis}
a=\frac{1}{\nu}\frac{\partial}{\partial y} u_x=\text{const}
\end{equation}
as the relevant nonequilibrium parameter. Obviously, $a_i=(\nu/\nu_i)a$.

Momentum transport is typically characterized by three rheological 
functions. The $xy$ element for shear stressess can be written in the form 
of a generalized Newton's viscosity law 
\begin{equation}
\label{3.13}
P_{xy}=-\eta_0 F_{\eta}(a) \frac{\partial u_x}{\partial y},
\end{equation} 
where 
\begin{equation}
\eta_0=nk_BT\sum_{i=1}^{N}\frac{x_i}{\nu_i}
\end{equation}
is the Navier-Stokes shear viscosity of the mixture \cite{GK56} and 
the explicit expression of $F_{\eta}(a)$ can be obtained from Eq.\ 
(\ref{3.9}) as
\begin{equation}
\label{3.14}
F_{\eta}(a)=\left(\sum_{i=1}^{N}\frac{x_i}{\nu_i}\right)^{-1}
\sum_{i=1}^{N}\frac{x_i\chi_i}{\nu_i^2}\sum_{j=1}^{N}\nu_{ij}
\chi_{ij}F_0(\gamma_{ij}).
\end{equation}
When $a=0$, $F_{\eta}=1$, so that this function describes the nonlinear 
rheological effects associated with the transport of the $x$ component of 
the momentum along the $y$ direction. Normal stresses are measured by the 
dimensionless viscometric functions $\Psi_{1,2}(a)$: 
\begin{equation}
\label{3.15}
\Psi_1(a)=\frac{P_{yy}-P_{xx}}{pa^2},
\end{equation} 
\begin{equation}
\label{3.16}
\Psi_2(a)=\frac{P_{zz}-P_{yy}}{pa^2},
\end{equation} 
where $p=nk_BT$. The explicit expressions for the viscometric functions can 
be easily obtained by substituting Eqs.\ (\ref{3.6})--(\ref{3.8}) into Eqs.\ 
(\ref{3.15}) and (\ref{3.16}). The result is
\begin{equation}
\label{3.17}
\Psi_1(a)=-\frac{2}{a^2}\sum_{i=1}^{N}\frac{x_i\chi_i}{\nu_i}
\sum_{j=1}^{N}\nu_{ij}
\chi_{ij}\gamma_{ij}\left[3F_1(\gamma_{ij})+
4F_2(\gamma_{ij})\right],
\end{equation} 
\begin{equation}
\label{3.18}
\Psi_2(a)=\frac{4}{a^2}\sum_{i=1}^{N}\frac{x_i\chi_i}{\nu_i}
\sum_{j=1}^{N}\nu_{ij}
\chi_{ij}\gamma_{ij}F_2(\gamma_{ij}).
\end{equation}

The heat flux introduces two relevant transport coefficients, $F_{\lambda}$ 
and $\Phi$, defined respectively as
\begin{equation}
\label{3.19}
q_y=-\lambda_0 F_{\lambda}(a) \frac{\partial T}{\partial y},
\end{equation}
\begin{equation}
\label{3.20}
q_x=-\lambda_0 \Phi(a) a \frac{\partial T}{\partial y}.
\end{equation}
Here, $\lambda_0$ is the Navier-Stokes thermal conductivity coefficient of 
the mixture given by \cite{GK56}
\begin{equation}
\label{3.21}
\lambda_0=\frac{5}{2}pk_B\sum_{i=1}^{N}\frac{x_i}{m_i\nu_i}.
\end{equation}
Equation (\ref{3.19}) can be interpreted as a generalization of Fourier's 
law with the thermal conductivity modified by the rheological quantity 
$F_{\lambda}(a)$, a nonlinear function of the shear rate and the parameters 
of the mixture. Equation (\ref{3.20}) provides information on the anisotropy 
induced by the Couette flow in the transport of energy. It gives a heat flux 
along the $x$ axis due to a thermal gradient along the $y$ axis. This effect 
is absent in the Navier-Stokes regime, being of first order in both the 
shear rate and the temperature gradient. Additional effects are described 
by the function $\Phi(a)$. The explicit expressions of $F_{\lambda}$ and 
$\Phi$ can be derived from Eqs.\ (\ref{3.10}) and (\ref{3.11}) as
\begin{eqnarray}
\label{3.22}
F_{\lambda}(a)&=&\left(\sum_{i=1}^{N}\frac{x_i}{m_i\nu_i}\right)^{-1} 
\sum_{i=1}^{N}\frac{x_i\chi_i^2}{5m_i\nu_i^2}\sum_{j=1}^{N}\nu_{ij}
\chi_{ij}\left\{3F_1(\gamma_{ij})+2F_2(\gamma_{ij})
\right. \nonumber\\
& & \left.
+2\frac{\nu^2 a^2}{\nu_i^2}
\left[4F_4(\gamma_{ij})+4F_3(\gamma_{ij})+
F_2(\gamma_{ij})\right]\right\},
\end{eqnarray} 
\begin{eqnarray}
\label{3.23}
\Phi(a)&=&-\frac{2}{5}\nu 
\left(\sum_{i=1}^{N}\frac{x_i}{m_i\nu_i}\right)^{-1} \sum_{i=1}^{N}
\frac{x_i\chi_i^2}{m_i\nu_i^3}\sum_{j=1}^{N}\nu_{ij}
\chi_{ij}\left\{5F_2(\gamma_{ij})+2F_3(\gamma_{ij})
\right.\nonumber\\
& & \left.
+2\frac{\nu^2 a^2}{\nu_i^2}\left[4F_5(\gamma_{ij})+
8F_4(\gamma_{ij})+5F_3(\gamma_{ij})+
F_2(\gamma_{ij})\right]\right\}.
\end{eqnarray}

Before closing this Section, let us explicitly write the corresponding 
expressions of the generalized transport coefficients in 
the simple case of particles mechanically equivalent. In this limit, 
$m_i=m$, $\kappa_{ij}=\kappa$, $\nu=A n(2\kappa/m)^{1/2}$, 
$\gamma_i=\gamma$, and $\chi_i=1$. Consequently, Eqs.\ 
(\ref{3.14}), (\ref{3.17}), (\ref{3.18}), (\ref{3.22}), and (\ref{3.23}) 
reduce, respectively, to  
\begin{equation}
\label{3.23n1}
F_{\eta}(a)=F_0(\gamma),
\end{equation}
\begin{equation}
\label{3.23n2}
\Psi_1(a)=-\frac{2}{a^2}\gamma\,\left[3F_1(\gamma)+4F_2(\gamma)\right],
\end{equation}
\begin{equation}
\label{3.23n3}
\Psi_2(a)=\frac{4}{a^2}\gamma\, F_2(\gamma),
\end{equation}
\begin{equation}
\label{3.23n4}
F_{\lambda}(a)=\frac{1}{5}\left\{3F_1(\gamma)+2F_2(\gamma)+2a^2\left[
4F_4(\gamma)+4F_3(\gamma)+F_2(\gamma)\right]\right\},
\end{equation}
\begin{equation}
\label{3.23n5}
\Phi(a)=-\frac{2}{5}\left\{5F_2(\gamma)+2F_3(\gamma)+2a^2\left[
4F_5(\gamma)+8F_4(\gamma)+5F_3(\gamma)+F_2(\gamma)\right]\right\}.
\end{equation}
All these expressions coincide with those previously derived from the BGK 
equation \cite{BSD87,MG98}.

\section{An illustrative example: A binary mixture}
\label{sec5}

Thus far, all the results hold for an $N$-component mixture. For the 
sake of illustration, in this Section we will consider the case of a binary 
mixture ($N=2$). Therefore, the corresponding set of two 
implicit equations (\ref{3.2}) give $\gamma_1$ and $\chi_1$ as functions 
of $a$ and the parameters of the mixture, namely, the mass ratio $\mu\equiv 
m_1/m_2$, the concentration ratio $\delta\equiv 
x_1/x_2$, and the size ratios $w_{11}\equiv\kappa_{11}/\kappa_{12}$, and 
$w_{22}\equiv\kappa_{22}/\kappa_{12}$. Although the natural nonequilibrium 
parameter is the shear rate $a$, from a practical point of view it is 
convenient to take $\gamma_1$ as an independent variable and express all the 
unknowns in terms of this parameter. In this sense, $\chi_1$ is defined by 
the implicit equation
\begin{equation}
\label{4.1}
\left(\frac{\nu_1}{\nu_2}\right)^2\frac{\sum_{i=1}^{2}
\nu_{1i}\chi_{1i}F_1(\gamma_{1i})}{\sum_{i=1}^{2}
\nu_{2i}\chi_{2i}F_1(\gamma_{2i})}
=\frac{3\nu_{12}(1-\chi_{12})
+2\sum_{i=1}^{2}\nu_{1i}\chi_{1i} 
\gamma_{1i}\left[ 2F_2(\gamma_{1i})+
3F_1(\gamma_{1i})\right]}{3\nu_{21}(1-\chi_{21})
+2\sum_{i=1}^{2}\nu_{2i}\chi_{2i} 
\gamma_{2i}\left[2F_2(\gamma_{2i})+
3F_1(\gamma_{2i})\right]},
\end{equation}
and the shear rate $a$ is given by 
\begin{equation}
\label{4.2}
a^2=\frac{1}{2}\left(\frac{\nu_1}{\nu}\right)^2\frac{3
\nu_{12}(1-\chi_{12})
+2\sum_{i=1}^{2}\nu_{1i}\chi_{1i} 
\gamma_{1i}\left[2F_2(\gamma_{1i})+
3F_1(\gamma_{1i})\right]}{\sum_{i=1}^{2}\nu_{1i}\chi_{1i}
F_1(\gamma_{1i})}.
\end{equation}
In the above equations, $\gamma_2$ and $\chi_2$ are 
related to $\gamma_1$ and $\chi_1$ through Eqs.\ (\ref{2.17}) and 
(\ref{3.5}). In general, Eq.\ (\ref{4.1}) must be solved numerically. 
However, $\gamma_1$ and $\chi_1$ can be explicitly derived in some limiting 
cases. For instance, in the limit of small shear rates
$\gamma_1\rightarrow \gamma_1^{(0)} a^2$, and 
$\chi_1\rightarrow 1+\chi_1^{(0)} a^2$, where 
\begin{equation}
\label{4.3}
\gamma_1^{(0)}=\frac{1}{5}\left(\frac{\nu}{\nu_1}\right)^2
\frac{\nu_1+\delta \nu_2}{\mu\nu_1+\delta \nu_2},
\end{equation}
\begin{equation}
\label{4.4}
\chi_1^{(0)}=\frac{1}{3}\frac{(\mu^2-1)(1+\mu)(1+\delta)}{\mu}
\frac{\nu_{12}}{\mu\nu_1+\delta \nu_2}.
\end{equation}

\begin{figure}
\begin{center}
\parbox{0.7\textwidth}{
\epsfxsize=\hsize \epsfbox{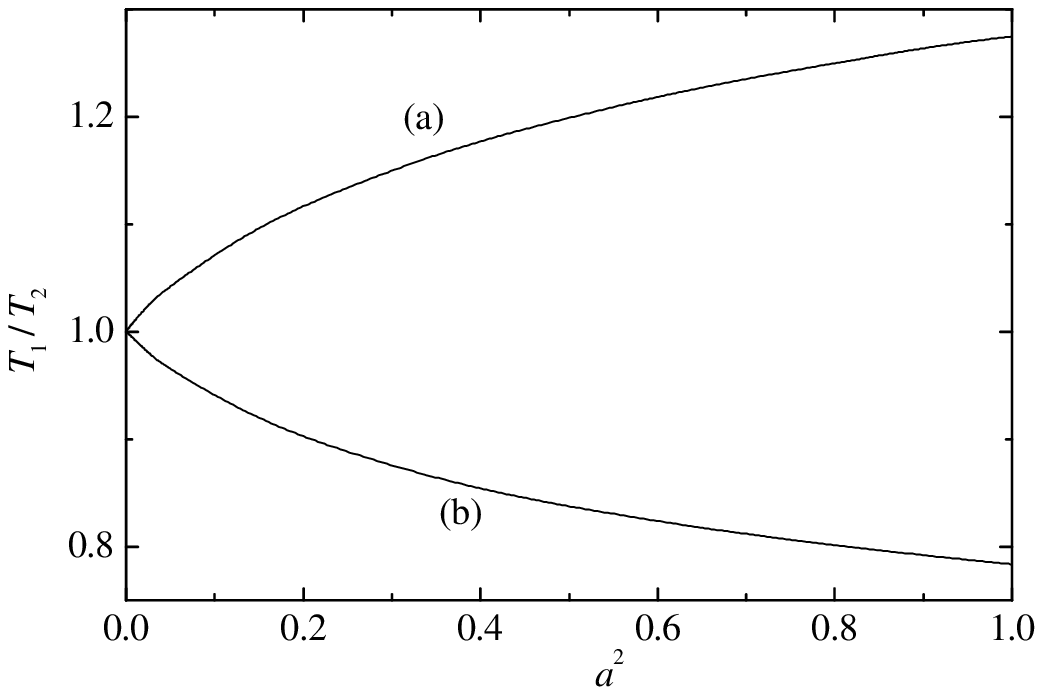}}
\caption{Shear-rate dependence of the temperature ratio $T_1/T_2$ for 
$w_{11}=w_{22}=1$, $\delta=2$, and 
two values of the mass ratio $\mu$: (a) $\mu=2$, and (b) $\mu=0.5$. 
\label{fig1}
}
\end{center}
\end{figure}

The temperature ratio $T_1/T_2$ is a measure of the lack of equidistribution 
of the kinetic energy. This quantity is plotted in Fig.\ \ref{fig1} as a 
function of the shear rate for $w_{11}=w_{22}=1$, a concentration ratio 
$\delta=2$ and two different values of the mass ratio: $\mu=0.5$ and 
$\mu=2$. We observe that the kinetic energy per particle of the solvent 
component $1$ is larger (smaller) than that of the solute component $2$ if 
the former is heavier (ligther) than the latter. The fact that $T_1\neq T_2$ 
for arbitrary shear rates indicates that the standard one-temperature 
Chapman-Enskog theory \cite{CC70} fails, even for not very disparate masses. 
In this case, the so-called two-fluid theory must be used 
\cite{two-temperature}. On the other hand, the monotonic behavior for 
$T_1/T_2$ found in the steady Couette flow contrasts with the one observed 
in the uniform shear flow \cite{MGS95}. In the latter problem, for given 
values of the parameters of the mixture, there exists a particular 
nonzero value of the shear rate $a$ such that $T_1=T_2$. 

\begin{figure}
\vspace{0.5cm}
\begin{center}
\parbox{0.7\textwidth}{
\epsfxsize=\hsize \epsfbox{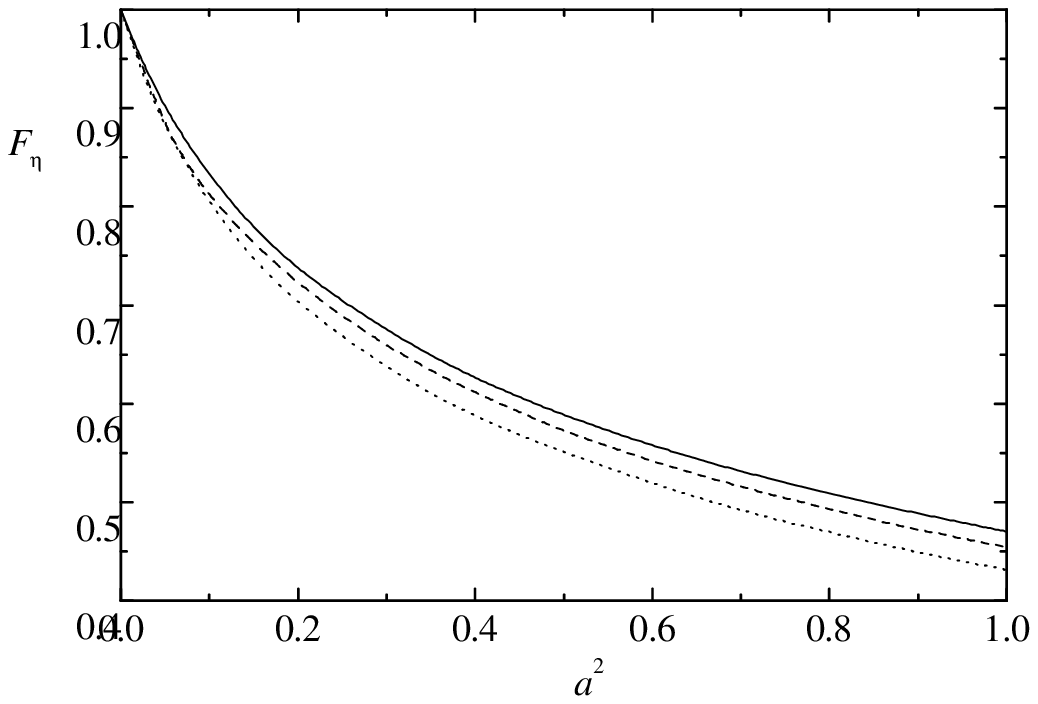}}
\caption{Shear-rate dependence of the viscosity function $F_{\eta}$ for 
$w_{11}=w_{22}=1$, $\delta=2$, and 
three values of the mass ratio $\mu$: $\mu=0.5$ (solid line),
$\mu=1$ (dashed line), and $\mu=2$ (dotted line). 
\label{fig2}}
\end{center}
\end{figure}

The viscosity function $F_{\eta}$ of a binary mixture under Couette flow 
is defined by Eq.\ (\ref{3.14}). For small shear rates this 
function behaves as $F_{\eta}\rightarrow 1+F_{\eta}^{(0)} a^2$, where 
\begin{eqnarray}
\label{4.5}
F_{\eta}^{(0)}&=&\left(\frac{x_1}{\nu_1}+\frac{x_2}{\nu_2}\right)^{-1}\left\{
\frac{x_1}{\nu_1}\left[M\chi_1^{(0)}\left(\frac{1}{M}-
2\frac{\nu_{12}}{\nu_1}(1+\delta)\right)-18\gamma_1^{(0)}\right] 
\right.\nonumber\\
& & 
\left. 
+\frac{x_2}{\nu_2}
\left[M\chi_1^{(0)}\left(2\frac{\nu_{21}}{\nu_2}(1+\delta)
-\frac{\delta}{M}\right)-18\frac{\mu \nu_1^2}{\nu_2^2}
\gamma_1^{(0)}\right] \right\},
\end{eqnarray}
with $M\equiv \mu/(1+\mu)^2$. Depending on the values of the parameters of 
the mixture, the super-Burnett contribution $F_{\eta}^{(0)}$ can be positive 
or negative, so for small rates the viscosity function can increase or 
decrease as the shear rate increases. Nevertheless, outside this small 
region, $F_{\eta}$ monotonically decreases as the shear rate increases, 
regardless of the values of the parameters of the mixture as shown in 
Fig.\ \ref{fig2}.

In addition, at a given value of the shear rate, the 
viscosity function increases as the mass ratio $m_1/m_2$ decreases when 
$n_1/n_2>1$. The viscometric functions $\Psi_{1,2}$ are given by Eqs.\ 
(\ref{3.17}) and (\ref{3.18}), respectively. In the limit $a\rightarrow 0$, 
one gets
\begin{equation} 
\label{4.6}
\Psi_1(0)=-14\left(x_1+x_2\mu 
\frac{\nu_1^2}{\nu_2^2} \right) \gamma_1^{(0)},
\end{equation}
\begin{equation} 
\label{4.7}
\Psi_2(0)=4\left(x_1+x_2\mu 
\frac{\nu_1^2}{\nu_2^2} \right) \gamma_1^{(0)}.
\end{equation}

\begin{figure}
\begin{center}
\parbox{0.7\textwidth}{
\epsfxsize=\hsize \epsfbox{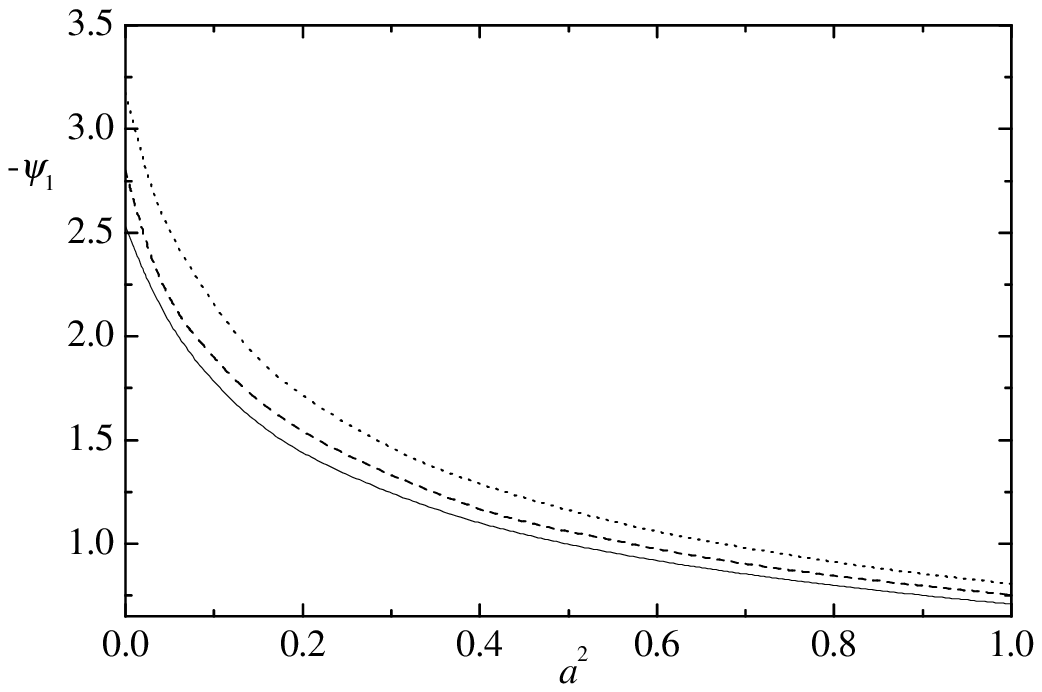}}
\caption{Shear-rate dependence of the first viscometric function $-\Psi_1$ 
for $w_{11}=w_{22}=1$, $\delta=2$, and 
three values of the mass ratio $\mu$: $\mu=0.5$ (solid line),
$\mu=1$ (dashed line), and $\mu=2$ (dotted line). 
\label{fig3}}
\end{center}
\end{figure}

\begin{figure}
\begin{center}
\parbox{0.7\textwidth}{
\epsfxsize=\hsize \epsfbox{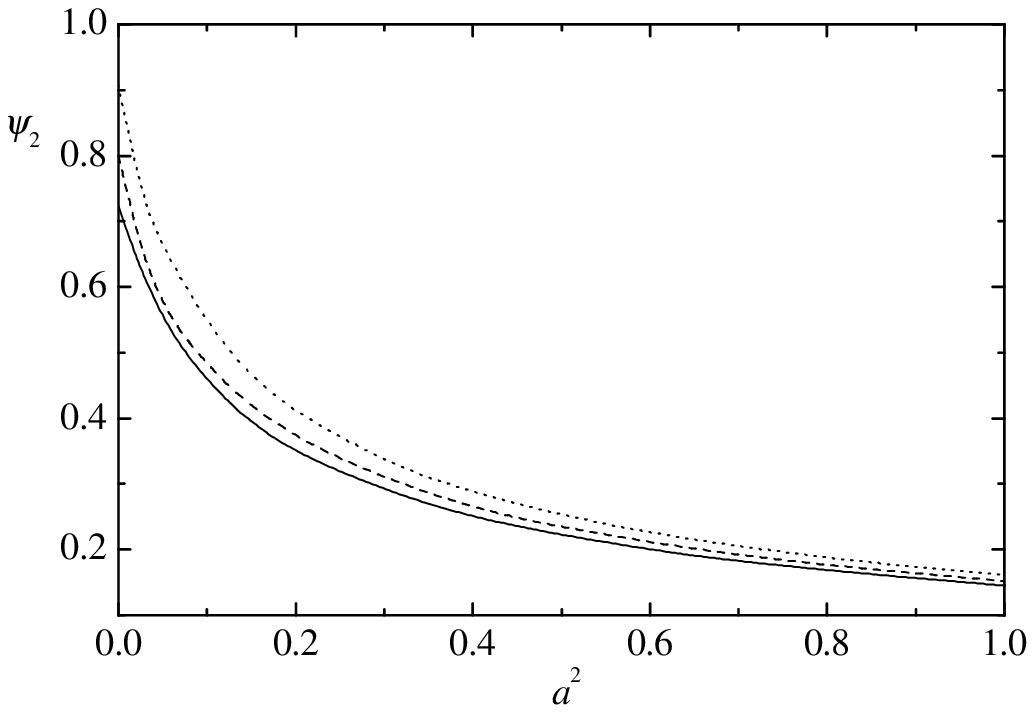}}
\caption{Shear-rate dependence of the second viscometric function $\Psi_2$ 
for $w_{11}=w_{22}=1$, $\delta=2$, and 
three values of the mass ratio $\mu$: $\mu=0.5$ (solid line),
$\mu=1$ (dashed line), and $\mu=2$ (dotted line). 
\label{fig4}}
\end{center}
\end{figure}

Notice that $\Psi_{1,2}(0)$ are Burnett transport coefficients. 
The shear rate dependence of the (dimensionless) viscometric functions 
$\Psi_{1,2}(a)$ is shown in Figs.\ \ref{fig3} and \ref{fig4}, 
respectively, for the same cases as in Fig.\ \ref{fig2}. The first 
viscometric function is always negative while the second one is positive. 
In general, the magnitude of both 
viscometric functions decreases with the shear rate. On the other hand, for 
zero shear rates,  $|\Psi_{1,2}(0)|_{\mu=0.5}<|\Psi_{1,2}(0)|_{\mu=1}<
|\Psi_{1,2}(0)|_{\mu=2}$. These inequalities are kept for finite shear 
rates. 

The transport of energy is characterized by the coefficients $F_{\lambda}$ 
and $\Phi$. For small shear rates, $F_{\lambda}\rightarrow 
1+F_{\lambda}^{(0)}a^2$ and $\Phi\rightarrow \Phi(0)$, with
\begin{eqnarray}
\label{4.8}
F_{\lambda}^{(0)}&=& 
\left(\frac{x_1}{m_1\nu_1}+\frac{x_2}{m_2\nu_2}\right)^{-1}\left\{
\frac{x_1}{m_1\nu_1}\left[2M\chi_1^{(0)}\left(\frac{1}{M}-
\frac{\nu_{12}}{\nu_1}(1+\delta)\right)-\frac{252}{5}\gamma_1^{(0)}-\frac{
36}{5}\frac{\nu^2}{\nu_1^2}\right] 
\right.\nonumber\\
& & 
\left. 
+\frac{x_2}{m_2\nu_2}
\left[2M\chi_1^{(0)}\left(\frac{\nu_{21}}{\nu_2}(1+\delta)
-\frac{\delta}{M}\right)-\frac{252}{5}\frac{\mu \nu_1^2}{\nu_2^2}
\gamma_1^{(0)}-\frac{36}{5}\frac{\nu^2}{\nu_2^2}\right] \right\}.
\end{eqnarray}
\begin{equation}
\label{4.9}
\Phi(0)=-\frac{14}{5}\frac{\nu}{\nu_1} \frac{\delta+\mu(\nu_1/\nu_2)^2}
{\delta+\mu(\nu_1/\nu_2)}.
\end{equation}

\begin{figure}
\begin{center}
\parbox{0.7\textwidth}{
\epsfxsize=\hsize \epsfbox{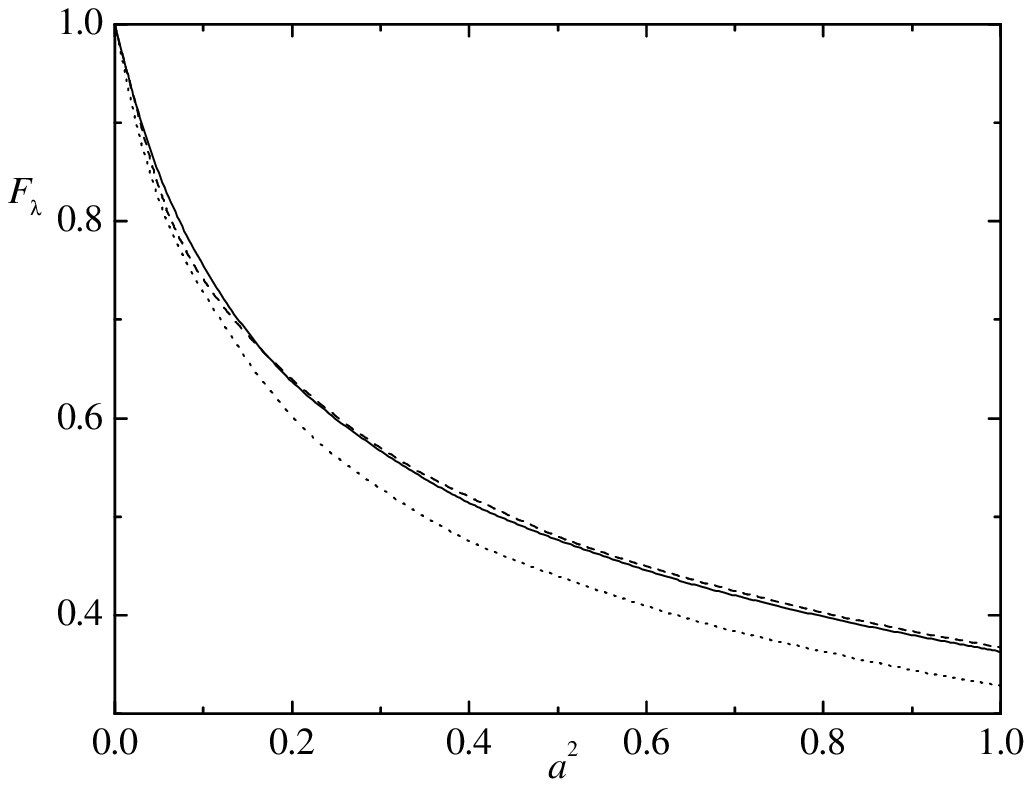}}
\caption{Shear-rate dependence of the generalized thermal 
conductivity function $F_{\lambda}$ for $w_{11}=w_{22}=1$, $\delta=2$, and 
three values of the mass ratio $\mu$: $\mu=0.5$ (solid line),
$\mu=1$ (dashed line), and $\mu=2$ (dotted line). 
\label{fig5}}
\end{center}
\end{figure} 

\begin{figure}
\begin{center}
\parbox{0.7\textwidth}{
\epsfxsize=\hsize \epsfbox{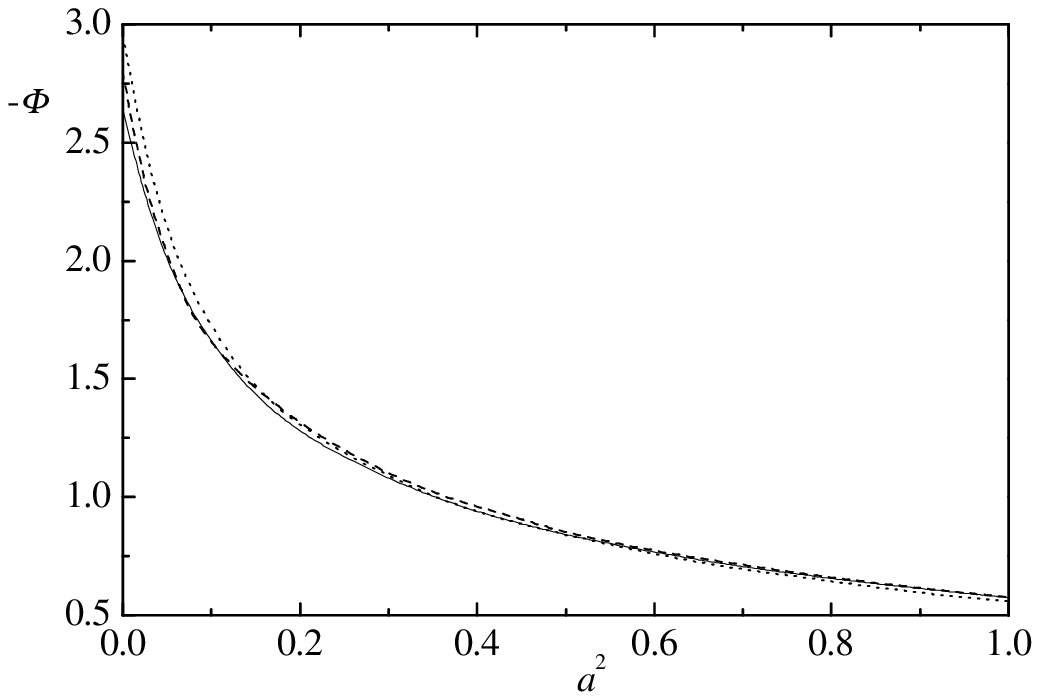}}
\caption{Shear-rate dependence of the cross coefficient 
$-\Phi$ for $w_{11}=w_{22}=1$, $\delta=2$, and 
three values of the mass ratio $\mu$: $\mu=0.5$ (solid line),
$\mu=1$ (dashed line), and $\mu=2$ (dotted line). 
\label{fig6}}
\end{center}
\end{figure}

It is apparent that the coefficient $F_{\lambda}^{(0)}$ presents a complex 
dependence on the parameters of the mixture. In general, and in accordance 
with Fig.\ \ref{fig5}, for finite values of the shear rate the 
generalized thermal conductivity coefficient $F_{\lambda}$ decreases as the 
shear rate increases with independence of the values of the parameters of 
the mixture. It is interesting to remark that 
the curves of $\mu=0.5$ and $\mu=1$ are practically the same in the range of 
shear rates considered. In 
Fig.\ \ref{fig6} we plot the shear rate dependence of the reduced 
coefficient $\Phi(a)$. We observe that $\Phi$ is negative and its 
magnitude decreases with the shear rate. Further, in the range of shear 
rates studied, this coefficient is practically independent of the mass 
ratio. This is rather surprising if one takes into account that this is a 
coefficient measuring complex coupling effects between both gradients and a 
more significant influence of the parameters of the mixture should be 
expected.  

Finally, let us note that in the tracer limit ($n_1/n_2\rightarrow 0$), all 
the results of the binary mixture reduce to those previously obtained from 
the GK model \cite{GS93}.

\section{Discussion}
\label{sec6}

In this paper we have considered a gas mixture of $N$ 
species in a steady nonequilibrium state involving combined heat and 
momentum transport. This state is usually referred to as the steady 
planar Couette flow. The system is a gas mixture enclosed between two 
relatively moving parallel plates which are kept in general at different 
temperatures. The shear rate and the thermal gradient are the two 
nonequilibrium parameters of the problem. Under these conditions, the 
density, flow velocity, and the temperature are nonhomogeneous, thus making 
the analysis of this problem more complicated than in other nonequilibrium 
situations. Further, the mixture is not restricted to specific values of the 
mass ratios, concentration ratios and size ratios. As a consequence, the 
transport properties of the system are {\em nonlinear} functions of the 
strengths of the gradients and of the parameters of the mixture. In order to 
offer a description as detailed as possible, we have used a kinetic model of 
the Boltzmann equation. Specifically, we have considered the GK model for 
mixtures, whose
reliability has been proved in the past in several nonequilibrium 
problems\cite{MMG96bis,GM98}. The explicit determination of the transport 
properties of the gas from an exact solution of the GK model has been the 
goal of this paper. It is important to remark that progress has been 
possible here due to the previous results derived in the limit cases of 
mechanically equivalent particles \cite{BSD87} and tracer particles 
\cite{GS93}.

Guided by the results obtained for a single gas\cite{BSD87}, we have found a 
consistent solution of the GK equation characterized by a uniform presure, 
and linear velocity and parabolic temperature profiles with respect to a 
certain scaled variable (distance times the local collision frequency). The 
consistency condition for the partial temperatures $T_i$ leads to a set of 
$N$ coupled transcendent equations, whose (numerical) solution gives $T_i$ 
in terms of the shear rate and the parameters of the mixture. Once the 
$T_i$'s are known, we explicitly evaluate the pressure tensor and the heat 
flux. The nonzero elements of the pressure tensor define three 
relevant transport coefficients: the generalized shear viscosity $F_{\eta}$ 
, Eqs.\ (\ref{3.13}) and (\ref{3.14}), the first 
viscometric function $\Psi_1$, Eqs.\ (\ref{3.15}) and (\ref{3.17}), and the 
second viscometric function $\Psi_{2}$, Eqs.\ (\ref{3.16}) and (\ref{3.18}).
While the viscosity function measures shear stresses, the two viscometric 
functions provide information on normal stresses. The heat flux turns out to 
be exactly proportional to the thermal gradient, with independence of its 
magnitude, but it defines two shear rate dependent transport coefficients:
the generalized thermal conductivity $F_{\lambda}$, Eqs.\ 
(\ref{3.19}) and (\ref{3.22}), and a generalized cross coefficient $\Phi$, 
Eqs.\ (\ref{3.20}) and (\ref{3.23}), which accounts for the transport of 
energy along the direction normal to the thermal gradient. For the sake of 
illustration, a binary mixture has been considered. The results show that, 
except perhaps in a narrow region small of shear rates, the magnitude of 
these coefficients decreases for finite shear fields. Consequently, the main 
effect of the shear flow is to inhibit the transport of momentum and energy 
across the system. In general, this inhibition is more significant when the 
excess component is heavier than the defect component in the case of the 
coefficients $F_{\eta}$ and $F_{\lambda}$, while the opposite happens in 
the case of the two viscometric functions. Concerning the cross 
coefficient $\Psi$ (which is negative), the results indicate that in the 
range of shear rates considered is practically independent of the values of 
the parameters of the mixture.

The problem studied here is worthwhile by itself. Furthermore, it can be 
taken as a starting point to analyze mutual diffusion under steady Couette 
flow. In the same way as done in the tracer limit \cite{GS95,G96}, the idea 
is to carry out a perturbation expansion in powers of the molar fraction 
gradient around the solution found in this paper. In the first order of the 
expansion, a generalized Fick's law is expected, where a shear rate 
dependent mutual diffusion tensor can be identified. 
On the other hand, it is apparent that the results reported here can also be 
of relevance in connection with computer simulations. In the context of 
molecular dynamics simulations, it is difficult to achieve 
large shear rates in the bulk region to clearly observe nonlinear effects in 
the transport of momentum and energy. One possibility to overcome such 
problems inherent to molecular dynamics in the low-density regime is to use 
the direct simulation Monte Carlo method \cite{B94}, which has been shown to 
be fruitful in several problems. We hope that the results derived here for 
the nonlinear transport coefficients may stimulate the performance of 
computer simulations to check the reliability of our predictions. Recent 
comparison made for a single gas under Couette flow supports the accuracy of 
the BGK results.  
 
\acknowledgments

I am grateful to Dr. A. Santos for a critical reading of the manuscript. 
Partial support from the DGES (Spain) through grant No. PB97-1501 and from 
the Junta of Extremadura (Fondo Social Europeo) through grant No. IPR98C019 
is acknowledged.

\appendix
\section{Velocity moments of the distribution functions}
\label{appA}

In this Appendix we evaluate the velocity integrals for the consistency 
conditions and for the momentum and heat fluxes. They correspond to the low 
order moments of the distribution functions $f_i$. Let us consider the 
formal solution to Eq.\ (\ref{2.12}) given by   
\begin{eqnarray}
\label{a1}
f_i&=&\left(1+\frac{v_y}{\nu_i}\frac{\partial}{\partial y}\right)^{-1}
\sum_{j=1}^{N}\frac{\nu_{ij}}{\nu_i}f_{ij}\nonumber\\
&=&\sum_{j=1}^{N}\frac{\nu_{ij}}{\nu_i}\Lambda_{ij} ,
\end{eqnarray}
where 
\begin{eqnarray}
\label{a2}
\Lambda_{ij}&=&\left(1+\frac{v_y}{\nu_i}\frac{\partial}
{\partial y}\right)^{-1} f_{ij} \nonumber\\
&=&
\sum_{k=0}^{\infty}(-v_y)^k\left(
\frac{1}{\nu_i}\frac{\partial}{\partial y}\right)^k\, f_{ij}.
\end{eqnarray}
We define the velocity integrals
\begin{equation}
\label{a3}
M_{k_1,k_2,k_3}^{ij}=\int d{\bf v}\, V_x^{k_1}V_y^{k_2}V_z^{k_3}\, 
\Lambda_{ij} .
\end{equation}
From a formal point of view, the moments defined in Eq.\ (\ref{a3}) can be 
directly obtained from the results derived in the one-component case 
\cite{BSD87,MG98} when one exploits the equivalence between the 
hydrodynamic profiles (\ref{2.14})--(\ref{2.16}) with those given in the 
single gas. Consequently, the moments (\ref{a3}) can be determined from 
comparison with those of the distribution function $f$ of the one-component 
gas by making the changes $\nu\rightarrow \nu_{ij}$, $\nu^{-1}\partial_y 
u_x\rightarrow a_i$, $T\rightarrow T_{ij}$, and 
$(\nu^{-1}\partial_y)^2T=-(2m/k_B)\gamma\rightarrow 
-(2m_i/k_B)\gamma_{ij}$. Taking into account this equivalence, we 
can explicitly write the first few moments of $\Lambda_{ij}$. Here, we 
display the moments with degree $k_1+k_2+k_3\leq 3$ appearing in the 
calculations performed along the main text. They are given by 
\cite{BSD87,MG98}
\begin{equation}
\label{a4}
M_{0,0,0}^{ij}=n_i,
\end{equation}
\begin{equation}
\label{a5}
M_{1,0,0}^{ij}=M_{0,1,0}^{ij}=M_{0,0,1}^{ij}=0,
\end{equation} 
\begin{equation}
\label{a6}
M_{2,0,0}^{ij}=\frac{n_ik_BT_{ij}}{m_i}\left[1+4\gamma_{ij}(
F_1+F_2)\right],
\end{equation} 
\begin{equation}
\label{a7}
M_{0,2,0}^{ij}=\frac{n_ik_BT_{ij}}{m_i}\left[1-2\gamma_{ij}(
F_1+2F_2)\right],
\end{equation} 
\begin{equation}
\label{a8}
M_{0,0,2}^{ij}=\frac{n_ik_BT_{ij}}{m_i}\left(1-2\gamma_{ij}F_1\right),
\end{equation} 
\begin{equation}
\label{a9}
M_{1,1,0}^{ij}=-\frac{n_ik_BT_{ij}}{m_i}F_0 a_i,
\end{equation} 
\begin{equation}
\label{a10}
M_{0,1,2}^{ij}=-\frac{n_ik_B^2T_{ij}}{m_i^2\nu_i}F_1
\frac{\partial}{\partial y}T_{ij},
\end{equation}
\begin{equation}
\label{a11}
M_{0,3,0}^{ij}=-\frac{n_ik_B^2T_{ij}}{m_i^2\nu_i}\left(2F_2
+F_1\right)\frac{\partial}{\partial y}T_{ij},
\end{equation}
\begin{equation}
\label{a12}
M_{1,2,0}^{ij}=2\frac{n_ik_B^2T_{ij}}{m_i^2\nu_i}\left(2F_3
+F_2\right)a_i\frac{\partial}{\partial y}T_{ij},
\end{equation}
\begin{equation}
\label{a13}
M_{1,0,2}^{ij}=2\frac{n_ik_B^2T_{ij}}{m_i^2\nu_i}F_2
a_i\frac{\partial}{\partial y}T_{ij},
\end{equation}
\begin{equation}
\label{a14}
M_{2,1,0}^{ij}=-\frac{n_ik_B^2T_{ij}}{m_i^2\nu_i}\left[F_1+2a_i^2
\left(4F_4+4F_3+F_2\right)\right]
\frac{\partial}{\partial y}T_{ij},
\end{equation} 
\begin{equation}
\label{a15}
M_{3,0,0}^{ij}=2\frac{n_ik_B^2T_{ij}}{m_i^2\nu_i}\left[3F_2
+2a_i^2\left(4F_5+8F_4+5F_3+F_2\right)\right]a_i
\frac{\partial}{\partial y}T_{ij}.
\end{equation} 
In these expressions, $\gamma_{ij}=\chi_{ij}\gamma_i$, 
$\chi_{ij}$  is defined in Eq.\ (\ref{3.3}) and 
\begin{equation}
\label{a16}
F_r\equiv F_r(\gamma_{ij})=\left(\frac{d}{d
\gamma_{ij}}\gamma_{ij}\right)^rF_0(
\gamma_{ij})
\end{equation}
with $F_0(\gamma_{ij})$ given by Eq.\ (\ref{3.4}).

The consistency conditions for the density and the flow velocity are 
easily seen to be verified by using Eqs.\ (\ref{a4}) and (\ref{a5}), 
respectively. The condition for the temperature on the other hand requires 
that 
\begin{equation}
\label{a17}
\sum_{j=1}^{N}\frac{\nu_{ij}}{\nu_i}\left(M_{2,0,0}^{ij}+M_{0,2,0}^{ij}+
M_{0,0,2}^{ij}\right)=3\frac{n_ik_BT_i}{m_i}.
\end{equation}
Substitution of Eqs.\ (\ref{a6})--(\ref{a8}) into Eqs.\ (\ref{a17}) yields 
the set of coupled equations (\ref{3.2}). The momentum and heat fluxes can 
be evaluated from the remaining velocity moments. The pressure tensor can be 
written as 
\begin{equation}
\label{a18}
{\sf P}=\sum_{i=1}^{N}\int d{\bf v} m_i {\bf V}{\bf V} 
f_i=\sum_{i=1}^{N}{\sf P}_i,
\end{equation}
where the nonzero elements of ${\sf P}_i$ are given by 
\begin{equation}
\label{a19}
\left\{P_{i,xx};P_{i,yy};P_{i,zz};P_{i,xy}\right\}=\sum_{j=1}^{N}
m_i\frac{\nu_{ij}}{\nu_i}\left\{M_{2,0,0}^{ij};M_{0,2,0}^{ij};M_{0,0,2}^{ij};
M_{1,1,0}^{ij}\right\}.
\end{equation}
The heat flux can be obtained in a similar way. It can be also written as 
\begin{equation}
\label{a20}
{\bf q}=\sum_{i=1}^{N}\int d{\bf v} \frac{m_i}{2} V^2 {\bf V} 
f_i=\sum_{i=1}^{N}{\bf q}_i,
\end{equation}
where 
\begin{equation}
\label{a21}
q_{y,i}=\sum_{j=1}^{N}\frac{m_i}{2}\frac{\nu_{ij}}{\nu_i}
\left(M_{2,1,0}^{ij}+M_{0,3,0}^{ij}+M_{0,1,2}^{ij}\right),
\end{equation}
\begin{equation}
\label{a22}
q_{x,i}=\sum_{j=1}^{N}\frac{m_i}{2}\frac{\nu_{ij}}{\nu_i}
\left(M_{3,0,0}^{ij}+M_{1,2,0}^{ij}+M_{1,0,2}^{ij}\right).
\end{equation}

The explicit expressions for the fluxes and transport coefficients 
appearing in the text can be easily derived from Eqs.\ 
(\ref{a18})--(\ref{a22}).


\begin{references}
\bibitem{M89} J. A. McLennan, Introduction to Nonequilibrium 
Statistical Mechanics, Prentice Hall, Englewood Cliffs, NJ, 1989.


\bibitem{CC70}S. Chapman, T. G. Cowling, The Mathematical 
Theory of Nonuniform Gases, Cambridge University Press, 
Cambridge, 1970.

\bibitem{TS95} M. Tij, A. Santos, Phys. Fluids 7 (1995) 2858.

\bibitem{RC97} D. Risso, P. Cordero, Phys. Rev. E 56 (1997) 489.

\bibitem{GK56} E. P. Gross, M. Krook, Phys. Rev. 102 (1956) 593.

\bibitem{BGK54}P.L. Bhatnagar, E. P. Gross, M. Krook, Phys. Rev. 94 
(1954) 511.

\bibitem{BSD87}J. J. Brey, A. Santos, J. W. Dufty, Phys. Rev. A  
36 (1987) 2842. 

\bibitem{KDSB89}C. S. Kim, J. W. Dufty, A. Santos, J. J. Brey, 
Phys. Rev. A 40 (1989) 7165.

\bibitem{GS93}V. Garz\'o and A. Santos, Phys. Rev. E 48 (1993) 256.

\bibitem{MGS94} C. Mar\'{\i}n, V. Garz\'o, A. Santos, J. Stat. Phys. 75 
(1994) 797.

\bibitem{MGS95} C. Mar\'{\i}n, V. Garz\'o, A. Santos, 
Phys. Rev. E 52 (1995) 3812.

\bibitem{MMG96} C. Mar\'{\i}n, J. J. Montanero, V. Garz\'o, Physica A 225 
(1996) 235.

\bibitem{Abramowitz}M. Abramowitz, I. A. Stegun, Handbook of 
Mathematical Functions, Dover, New York, 1972.


\bibitem{MG98} J. M. Montanero, V. Garz\'o, Phys. Rev. E 58 (1998) 1836.

\bibitem{two-temperature}C. J. Goebel, S. M. Harris, E. A. Johnson, 
Phys. Fluids 19 (1976) 627.

\bibitem{MMG96bis}C. Mar\'{\i}n, J. M. Montanero, V. Garz\'o, Mol. Phys. 88 
(1996) 1249.

\bibitem{GM98} V. Garz\'o, C. Mar\'{\i}n, in: S. G. Pandalai (Ed.), 
Recent Research Developments in Physics of Fluids 1, Transworld Research 
Network, Trivandrum, 1998.

\bibitem{GS95}V. Garz\'o, A. Santos, Phys. Rev. E 52 (1995) 4942.
\bibitem{G96} V. Garz\'o, Physica A 234 (1996) 108.
\bibitem{B94} G. A. Bird, Molecular Gas Dynamics and the Direct 
Simulation Monte Carlo of Gas Flows, Clarendon, Oxford, 1994.


\end{references}
\end{document}